\documentclass[prd,superscriptaddress,twocolumn,epsf,nofootinbib]{revtex4-1}

\usepackage{amsfonts,amssymb,amsmath} 
\usepackage{mathrsfs} 
\usepackage{color} 
\usepackage{graphicx}
\usepackage{dcolumn}
\usepackage{bm}
\usepackage{siunitx}
 
\newcommand{\B}[1]{{\mathbb #1}} 
\newcommand{\C}[1]{{\mathcal #1}}

\newcommand{\BS}[1]{{\boldsymbol #1}} 
 
\newcommand{\nn}{\nonumber}

\newcommand{\half}{\frac 12}

\newcommand{\Slash}[1]{{\ooalign{\hfil#1\hfil\crcr\raise.167ex\hbox{/}}}}

\begin{document} 
\title{eV-scale sterile neutrinos from an extra dimension}
\author{Shinsuke Kawai} 
\email{kawai@skku.edu} 
\affiliation{
	Department of Physics, 
	Sungkyunkwan University, 
	Suwon 16419, Republic of Korea}
\author{Nobuchika Okada}
\email{okadan@ua.edu}
\affiliation{
	Department of Physics and Astronomy, 
	University of Alabama, 
	Tuscaloosa, AL35487, USA} 
\date{\today} 
 
\begin{abstract}
Motivated by the short-baseline neutrino oscillation anomalies that suggest the existence of sterile neutrinos at the eV scale, we construct a scenario of a seesaw mechanism for 3+1 light neutrinos implemented by warped compactification of an extra dimension.
As the seesaw mechanism necessitates at least two right-handed neutrinos at mass scales much larger than eV, incorporating an eV-scale sterile neutrino into a seesaw entails large mass hierarchies among the singlet neutrinos.
We show that such hierarchies can be naturally explained by moderate fluctuations of the five-dimensional fermion mass parameters.
\end{abstract}
 
\maketitle 

\section{Introduction}

The progress of neutrino physics is remarkable.
Today, precise measurements of the neutrino mass-squared differences and the mixing angles are becoming available \cite{Tanabashi:2018oca}.
The existence of light sterile neutrinos, the (normal or inverted) mass hierarchy of the active neutrinos, and $CP$ violation in the lepton sector are now major questions that are expected to be settled by experiments in the immediate future. 
In particular, the MiniBooNE Collaboration \cite{Aguilar-Arevalo:2018gpe} recently reported 
4.8 $\sigma$ excess in the $\nu_e$ and $\overline{\nu_e}$ appearance experiments, and, combined with the Liquid Scintillator Neutrino Detector (LSND) results, the confidence level reaches 6.0 $\sigma$.
Although consistency with other experiments is still a matter of debate, if neutrino oscillations are to be responsible for the anomalies\footnote{A possible nuclear physics origin of the anomalies has also been suggested.
See Refs. \cite{Giunti:2019aiy,Diaz:2019fwt} for recent reviews.}
the combined LSND and MiniBooNE results strongly signal the existence of sterile neutrinos around the mass scale of eV.
If confirmed, the discovery of sterile neutrinos would have a huge impact on particle physics and cosmology, not merely on the physics of neutrinos.

A theoretical challenge would then to work out how the eV-scale sterile neutrinos may be incorporated in a sensible theory beyond the Standard Model (SM).
Since these leptons are SM gauge singlets, and as the seesaw mechanism is one of the leading candidates for the generation mechanism of small nonzero neutrino masses \cite{Minkowski:1977sc,Yanagida:1979as,GellMann:1980vs,Mohapatra:1979ia}, it seems reasonable to discuss a model in the framework of a type I seesaw mechanism.
For a successful seesaw, at least two heavy singlet (right-handed) neutrinos are necessary.
It is also well known that the $U(1)_{\rm B-L}$-extended Standard Model, which is obtained by gauging the global $U(1)_{\rm B-L}$ symmetry of the SM and gives rise to the type I seesaw when the $U(1)_{\rm B-L}$ symmetry is spontaneously broken, is anomaly-free only if there are three right-handed neutrinos.
Thus, the minimal anomaly-free model must have one light (eV-scale) and two heavy (seesaw-scale) right-handed neutrinos, embedded in the $U(1)_{\rm B-L}$-extended Standard Model.
An obvious question then is how such large mass hierarchies may be realized in the neutrino sector.

Let us recall the ordinary type I seesaw mechanism with three heavy right-handed neutrinos with the mass matrix (in the order of the three left-handed and three right-handed neutrinos)
\begin{align}
\left(
\begin{matrix}
  O & m_D^{3\times 3} \\
  &\\
  (m_D^{3\times 3})^T & M^{3\times 3}
\end{matrix}
\right).
\end{align}
On diagonalizing the matrix (or integrating out the right-handed neutrinos), the $3\times 3$ left-handed neutrino mass matrix is given by the seesaw formula
\begin{align}
  m_\nu^{3\times 3}\approx m_D^{3\times 3}(M^{3\times 3})^{-1}(m_D^{3\times 3})^T.
\end{align}
For the Dirac Yukawa couplings of ${\C O}(1)$, $m_D^{3\times 3}$ is of the order of the weak scale. 
Thus, for the neutrino masses of $m_\nu^{3\times 3}\lesssim$ eV, the seesaw scale turns out to be $M^{3\times 3}\sim 10^{13}$ GeV.
In our $3+1$ $(+2)$ neutrino model, in which one of the right-handed neutrinos is at the eV scale and the other two are much heavier, the mass matrix is written
\begin{align}\label{eqn:312MM}
\left(
\begin{matrix}
  M_\nu^{4\times 4} & m_D^{4\times 2} \\
  &\\
  (m_D^{4\times 2})^T & M^{2\times 2}
\end{matrix}
\right),
\end{align}
where
\begin{align}\label{eqn:Mnu44}
  M_\nu^{4\times 4}=
\left(
  \begin{matrix}
  & & & *\\
  & O & & *\\
  & & & * \\
  * & * & * & M_1
  \end{matrix}
\right)
\end{align}
in which the asterisks are nonzero matrix elements and $M_1 \sim $ eV.
The seesaw formula for the 3+1 neutrino masses is then
\begin{align}
  m_\nu^{4\times 4}
  \approx M_\nu^{4\times 4}-m_D^{4\times 2}(M^{2\times 2})^{-1}(m_D^{4\times 2})^T.
\end{align}
For a successful seesaw mechanism with $m_\nu^{4\times 4}\lesssim$ eV, the entries of the matrix elements need to be severely tuned; the nondiagonal elements of Eq. \eqref{eqn:Mnu44} need to be $\lesssim $ eV, and
$M_\nu^{4\times 4}\ll m_D^{4\times 2}\ll M^{2\times 2}$.
In this paper, we propose a simple compactification scenario that gives rise to a naturally tuned desired neutrino mass matrix \eqref{eqn:312MM}.

\section{The model}

We consider the minimal $U(1)_{\rm B-L}$-extended Standard Model embedded in the Poincar\'e patch of a five-dimensional anti-de Sitter (AdS) spacetime, the so-called Randall-Sundrum background \cite{Randall:1999ee}; see also
\cite{Gogberashvili:1998iu,Gogberashvili:1998vx,Gogberashvili:1999tb}.
The metric is
\begin{align}\label{eqn:Poincare}
	ds^2
	=&\; g_{\mu\nu}dx^\mu dx^\nu\nn\\
	=&\; e^{-2\sigma(y)}(-dt^2+dx_1^2+dx_2^2+dx_3^2)+dy^2,
\end{align}
where $\sigma(y)=ky$ 
and $k$ is the inverse of the AdS radius which is related to the five-dimensional AdS cosmological constant by $\Lambda=-6k^2$.
The fifth dimension coordinate $y$ is in the segment $0\leq y\leq \pi r_c$, which is considered as a circle of radius $r_c$,
\begin{align}
	y=r_c\phi,\quad
	\phi\simeq \phi+2\pi	
\end{align}
subject to ${\B Z}_2$ orbifold projection
	$\phi\simeq -\phi$.
The compactification radius is taken to be $\pi r_c k\sim {\C O}(10)$.
The SM fields are assumed to be localized on the hypersurface at the orbifold fixed point $y=\pi r_c$ (``the IR brane"), whereas the $U(1)_{\rm B-L}$ gauge field and the right-handed neutrino fields extend over the fifth dimension (``the bulk").
The $U(1)_{\rm B-L}$ Higgs $\Phi$ resides only on the IR brane and has a vacuum expectation value $\langle\Phi\rangle=v_{\rm BL}$.
On the hypersurface at the other orbifold fixed point $y=0$ (``the UV brane"), we assume no matter field except a cosmological constant.
Because of the warped geometry, the four-dimensional (reduced) Planck mass $M_4=2.44\times 10^{18}$ GeV is related to the five-dimensional reduced Planck mass $M_5$ by
	$M_4^2=(1-e^{-2\pi r_c k})M_5^3/k\sim M_5^3/k$.
The matter contents of the lepton sector on the brane are the lepton doublets $\ell^\alpha$, the lepton singlets (sterile neutrinos) $N_i^c$, the SM Higgs doublet $h$, and the $U(1)_{\rm B-L}$ Higgs boson $\Phi$.
We use the convention in which the four-dimensional fermions are written in terms of the left-handed fields.
The properties of these fields are summarized in the following table:

\begin{center}
\begin{tabular}{c|cccc}\label{table:matter}
 Fields & $SU(2)_L$ & $U(1)_Y$ & $U(1)_{\rm B-L}$ & Location\\
  \hline\\
  $\ell^\alpha$ & ${\BS 2}$ & $-\half$ & $-1$ & IR brane\\
   $N^c_i$ & ${\BS 1}$ & 0 & $+1$ & Bulk \\
 $h$ & ${\BS 2}$ & $+\half$ & 0 & IR brane \\
 $\Phi$ & ${\BS 1}$ & 0 & $-2$ & IR brane
\end{tabular}
\end{center}

\subsection{Singlet neutrinos from bulk fermions}

The seesaw mechanism in the Randall-Sundrum background has been studied from various viewpoints \cite{Huber:2003sf,Perez:2008ee,Csaki:2008qq,Fong:2011xh,Iyer:2013hca}.
Our concern in this paper is to accommodate an eV-scale sterile neutrino in a seesaw by realizing large mass hierarchies among the right-handed neutrinos.\footnote{
Our model is analogous in spirit to the split seesaw scenario \cite{Kusenko:2010ik}
that realizes keV-scale sterile neutrino dark matter as domain wall fermions.
}
In our model, the sterile neutrinos $N_i^c$ are realized as the Kaluza-Klein (KK) zero modes of the bulk fermions.
The four-dimensional chiral fermions are identified as the zero modes of the five-dimensional Dirac fermions subject to the ${\B Z}_2$ orbifold projection \cite{Grossman:1999ra,Chang:1999nh,Davoudiasl:2000wi}.
We start with the five-dimensional action of the bulk fermions:
\begin{align}\label{eqn:Sbf}
	S_{\rm bf}=\int d^5x\sqrt{-g}\Big\{
	e^\mu_a\left(\frac i2 \overline\Psi\gamma^a D_\mu\Psi +\text{H.c.}\right)
	-{\rm sgn}(\phi)m\overline\Psi\Psi
	\Big\},
\end{align}
where the terms involving spin connections are omitted as they play no r\^ole in the following discussions.
The ${\rm sgn}(\phi)$ in the mass term is to keep this term even under the ${\B Z}_2$ parity of the orbifold.
The pentads are 
$e^a_\mu=e^{-\sigma}\delta^a_\mu$ for $a=0, 1, 2, 3$, $e^5_\mu=\delta^5_\mu$, and their inverse are 
$e_a^\mu=e^\sigma\delta_a^\mu$ for $a=0, 1, 2, 3$, $e_5^\mu=\delta_5^\mu$.
We use the four-dimensional representations of the five-dimensional gamma matrices
$\gamma^a=(\gamma^0,\gamma^1,\gamma^2,\gamma^3,i\gamma_5)$.
The chiral components of the bulk fermion fields are
\begin{align}
	\Psi_{\rm L,R}=\half (1\mp\gamma_5)\Psi,
\end{align}
which are decomposed into the KK modes:
\begin{align}
	\Psi_{\rm L,R}=\frac{1}{\sqrt{r_c}}\sum_{n=0}^\infty\psi_{\rm L,R}^{(n)}(x)e^{2\sigma}\widehat f_{\rm L,R}^{(n)}(y).
\end{align}
For a consistent KK reduction, the action \eqref{eqn:Sbf} should give the four-dimensional action for canonically normalized massive fermions $\psi_{\rm L,R}^{(n)}(x)$.
This is accomplished by requiring the orthonormality 
\begin{align}\label{eqn:orthonormal}
	\int d\phi\; e^\sigma\widehat f_{\rm L}^{(m)*}\widehat f_{\rm L}^{(n)}
	=\delta^{mn}
	=\int d\phi\; e^\sigma\widehat f_{\rm R}^{(m)*}\widehat f_{\rm R}^{(n)}
\end{align}
and the bulk equations
\begin{align}\label{eqn:EoM}
	\Big(\pm\frac{1}{r_c}\partial_\phi+m\Big)\widehat f_{\rm R,L}^{(n)}=m_ne^\sigma\widehat f_{\rm L,R}^{(n)}
\end{align}
for the KK mode functions $\widehat f_{\rm L,R}^{(n)}$.
Here $m_n$ are the KK masses.
The left modes $\widehat f_{\rm L}^{(n)}$ and the right modes $\widehat f_{\rm R}^{(n)}$ have opposite ${\B Z}_2$ parities.
The left and right nonzero modes are coupled by Eq. \eqref{eqn:EoM}, satisfy second-order differential equations, and are solved as Bessel functions.
We shall not discuss nonzero modes in this paper, as they do not give nontrivial physics at low energy.\footnote{
The nonzero modes give flavor-changing operators \cite{Kitano:2000wr}, but the constraints from the flavor-changing neutral currents are trivially satisfied for our parameter range.
}
For the zero modes, either $\widehat f_{\rm L}^{(0)}$ or $\widehat f_{\rm R}^{(0)}$ is nontrivial, since the ${\B Z}_2$ odd part is projected away.
Without losing generality, we shall choose the left component to be ${\B Z}_2$ even. 
This conforms to our convention that four-dimensional chiral fermions are expressed in terms of the left-handed fields.
The zero modes also satisfy Eq. \eqref{eqn:EoM} with a vanishing KK mass and are normalized by Eq. \eqref{eqn:orthonormal}.
Thus,
\begin{align}
\widehat f_{\rm L}^{(0)}
=\left(\frac{(1+2\nu)kr_c}{e^{(1+2\nu)\pi r_ck}-1}\right)^{1/2}
e^{\nu\sigma},\qquad
\widehat f_{\rm R}^{(0)}=0,
\end{align}
where $\nu=m/k$.
In our model\footnote{
The ${\B Z}_2$ parity anomaly cancels only for an even number of fermions in the bulk.
In our model, the SM fields are well localized on the brane but are of five-dimensional origin, so there are 16 bulk fermions per generation and 48 in total.
Therefore this model is free from both the ${\B Z}_2$ parity anomaly and the $U(1)_{\rm B-L}$ gauge anomaly.
} we consider three generations of five-dimensional fields
$\Psi_i$, $i=1,2,3$, with possibly different five-dimensional masses $m_i$.
Their left-handed parts have KK decomposition:
\begin{align}\label{eqn:Lmode}
	\Psi_{i,\rm L}
	=&\; \half (1-\gamma_5)\Psi_i\\
	=&\; \psi_{i,\rm L}^{(0)}(x)\left(
	\frac{(1+2\nu_i)k}{e^{(1+2\nu_i)\pi r_ck}-1}\right)^{1/2}
e^{(2+\nu_i)\sigma}+\cdots,\nn
\end{align}
where $\nu_i=m_i/k$.
The canonically normalized left-handed zero mode fields $\psi_{i,\rm L}^{(0)}(x)$ are identified as (the conjugate of) the right-handed neutrino fields $N_i^c$.
The five-dimensional mass parameters $m_i$ are naturally in the same order as the curvature scale $k$, but their signs can be positive or negative.
We shall assume one of them, say, $m_1$, is smaller than $-\half k$ and the other two are larger, namely,
\begin{align}\label{eqn:nues}
  1+2\nu_1<0,\quad 1+2\nu_2>0,\quad 1+2\nu_3>0.
\end{align}
These mild assumptions lead to the desired mass hierarchies of the singlet neutrinos, as shown below.

\subsection{Majorana masses}
The action for the $U(1)_{\rm B-L}$ Higgs field on the brane can be written
\begin{align}\label{eqn:SPhi}
  S_{\Phi}=&\int d^4x\int dy\sqrt{-g}\;\delta(y-\pi r_c)\nn\\
  &\times\left\{
  -g^{\mu\nu}(D_\mu\widetilde\Phi)^\dag (D_\nu\widetilde\Phi)
  -\lambda\Big(\widetilde\Phi^\dag\widetilde\Phi-\half\widetilde v_{\rm BL}^2\Big)^2\right\}.
\end{align}
Since $\delta(y-\pi r_c)$ is an operator of mass dimension one, $\widetilde\Phi$ has the same mass dimension (one) as a four-dimensional scalar.
By rescaling
\begin{align}\label{eqn:rescalePhi}
	\Phi=e^{-\pi r_c k}\widetilde\Phi,
	\quad 
	v_{\rm BL}=e^{-\pi r_c k}\widetilde v_{\rm BL},
\end{align}
the action \eqref{eqn:SPhi} reduces to that of the canonically normalized four-dimensional scalar $\Phi$ with spontaneously broken $U(1)_{\rm B-L}$ symmetry.
The exponential suppression of the four-dimensional breaking scale $v_{\rm BL}$ is a well-known feature of the Randall-Sundrum-type scenario \cite{Randall:1999ee}.

The Majorana masses of the singlet neutrinos are generated by the $U(1)_{\rm B-L}$ Higgs expectation value on the brane.
Let us consider the Majorana Yukawa term
\begin{align}
  S_{\rm MY}\label{eqn:SMY1}
  =-\int d^4x\int dy\sqrt{-g}\delta(y-\pi r_c)\frac{\lambda_{ij}}{M_5}
  \widetilde\Phi \overline\Psi_i\Psi_j,
\end{align}
where we have divided by $M_5\approx k$ to make $\lambda_{ij}$ dimensionless.
Upon KK reduction \eqref{eqn:Lmode} and rescaling of the field \eqref{eqn:rescalePhi}, we may write
\begin{align}
  S_{\rm MY}\label{eqn:SMY2}
  =-\int d^4x \lambda_{ij}^{\rm eff} \Phi\overline N_i^c N_j^c+\text{KK modes}.
\end{align}
The four-dimensional effective Majorana Yukawa is
\begin{align}
  \lambda_{ij}^{\rm eff}=&\;\sqrt{\frac{(1+2\nu_i)(1+2\nu_j)}{(e^{(1+2\nu_i)\pi r_c k}-1)(e^{(1+2\nu_j)\pi r_c k}-1)}}\crcr
  &\times e^{(1+\nu_i+\nu_j)\pi r_c k}\lambda_{ij},
\end{align}
which, introducing 
\begin{align}\label{eqn:omegai}
	\omega_i\equiv\sqrt{\frac{(1+2\nu_i)e^{(1+2\nu_i)\pi r_c k}}{e^{(1+2\nu_i)\pi r_c k}-1}},
\end{align}
is compactly written 
\begin{align}
  \lambda_{ij}^{\rm eff}=\omega_i\omega_j\lambda_{ij}
\end{align}
(no summation).
For $\pi k r_c\sim{\C O}(10)$, the assumptions \eqref{eqn:nues} lead to
$\omega_1\equiv\epsilon\ll 1$, $\omega_2\sim\omega_3\sim {\C O}(1)$.
Therefore, the four-dimensional Majorana Yukawa exhibits hierarchical structure:
\begin{align}\label{eqn:MYukawa}
  \lambda_{11}^{\rm eff}\sim &\; \epsilon^2\; \lambda_{11},\crcr
  \lambda_{1j}^{\rm eff}\sim &\; \epsilon\; \lambda_{1j},\quad j=(2,3),\crcr
  \lambda_{ij}^{\rm eff}\sim &\; \lambda_{ij},\quad i,j=(2,3). 
\end{align}

\subsection{Dirac masses}
The Dirac masses of the singlet neutrinos originate from the five-dimensional Dirac Yukawa term
\begin{align}\label{eqn:SDY1}
	S_{\rm DY}
	=-\int d^4x\int dy\sqrt{-g}\;\delta(y-\pi r_c)\Big(
	\frac{y_{\alpha i}}{\sqrt{M_5}}\overline L^\alpha H\Psi_i+\text{H.c.}\Big),
\end{align}
in which the factor of $1/\sqrt{M_5}\approx 1/\sqrt k$ has been introduced to make $y_{\alpha i}$ dimensionless.
The fields $H$ and $L^\alpha$ are the five-dimensional scalar and fermions that are related by rescaling to the Higgs $h$ and the lepton doublets $\ell^\alpha$ canonically normalized in four-dimensions:
\begin{align}\label{eqn:rescaleHL}
  h=e^{-\pi r_c k}H,
  \quad
  \ell^\alpha=e^{-\frac 32 \pi r_c k}L^\alpha.
\end{align}
Upon KK reduction \eqref{eqn:Lmode} and rescaling of the fields \eqref{eqn:rescaleHL}, the Dirac Yukawa term becomes
\begin{align}\label{eqn:SDY2}
    S_{\rm DY}
	=&-\int d^4x\;\Big(y_{\alpha i}^{\rm eff}\overline \ell^\alpha h N_i^c+\text{H.c.}\Big)+\text{KK modes},
\end{align}
with the effective Yukawa coupling
\begin{align}
	y_{\alpha i}^{\rm eff}
	=&\;\sqrt{
	\frac{1+2\nu_i}{e^{(1+2\nu_i)\pi r_ck}-1}}\;
    e^{(\half+\nu_i)\pi r_c k}y_{\alpha i}\crcr
    =&\;\omega_i\; y_{\alpha i}.
\end{align}
Thus, with our assumptions on the 5 dimensional mass parameters \eqref{eqn:nues} the effective Dirac Yukawa coupling are
\begin{align}\label{eqn:DYukawa}
  y_{\alpha 1}^{\rm eff}\sim &\;\epsilon\; y_{\alpha 1},\crcr
  y_{\alpha j}^{\rm eff}\sim &\; y_{\alpha j},\quad j=(2,3).
\end{align}

\section{Benchmark for the seesaw mechanism}

We have seen that the minimal $U(1)_{\rm B-L}$-extended Standard Model embedded in the warped five-dimensional background gives the Majorana Yukawa term \eqref{eqn:SMY2} and Dirac Yukawa term \eqref{eqn:SDY2} in the effective theory on the IR brane.
After the $U(1)_{\rm B-L}$ and the electroweak symmetry are spontaneously broken, the mass term of the neutrino sector becomes
\begin{align}
	S_{\nu}=-\int d^4x\left\{
	y_{\alpha i}^{\rm eff}\overline\ell^\alpha\langle h\rangle N^c_i+\text{H.c.}
	+\lambda_{ij}^{\rm eff}v_{\rm BL}\overline N^c_i N^c_j
	\right\}.
\end{align}
With the assumptions on the mass parameters \eqref{eqn:nues}, the Yukawa couplings exhibit the hierarchical structure \eqref{eqn:MYukawa} and \eqref{eqn:DYukawa}.
Let us now focus on the seesaw mechanism realized in this model.

Introducing
\begin{align}
	m_{\alpha i}\equiv y_{\alpha i}^{\rm eff}\langle h\rangle,
	\quad
	M_{ij}\equiv\lambda_{ij}^{\rm eff}v_{\rm BL},
\end{align}
the mass matrix reads
\begin{align}\label{eqn:MM}
\left(
\begin{array}{cccccc}
 0 & 0 & 0 & m_{11} & m_{12} & m_{13}\\
 0 & 0 & 0 & m_{21} & m_{22} & m_{23}\\
 0 & 0 & 0 & m_{31} & m_{32} & m_{33}\\
 m_{11} & m_{21} & m_{31} & M_{11} & M_{12} & M_{13}\\
 m_{12} & m_{22} & m_{32} & M_{21} & M_{22} & M_{23}\\
 m_{13} & m_{23} & m_{33} & M_{31} & M_{32} & M_{33}
\end{array}
\right).
\end{align}
The Higgs vacuum expectation value is $\langle h\rangle\approx 100$ GeV and the original Yukawa couplings $\lambda_{ij}$ and $y_{\alpha i}$ are all assumed to be ${\C O}(1)$.
We wish to realize the lightest sterile neutrino at the eV scale:
\begin{align}\label{eqn:M11}
  M_{11}= \lambda_{11}^{\rm eff} v_{\rm BL}\sim\epsilon^2 v_{\rm BL}\sim 1\;\text{eV}.
\end{align}
Next, the Dirac masses of the $m_{\alpha 1}$ components must be eV scale, 
$m_{\alpha 1}=y_{\alpha 1}^{\rm eff}\langle h\rangle\sim\epsilon\times 100\;\text{GeV}\sim 1\;\text{eV}$.
Hence, $\epsilon\sim 10^{-11}$, and then, from Eq. \eqref{eqn:M11},
\begin{align}
  v_{\rm BL}\sim 10^{13}\;\text{GeV}.
\end{align}
The other Dirac mass components are 
$m_{\alpha 2}=y_{\alpha 2}^{\rm eff}\langle h\rangle\sim 100\; \text{GeV}$
and
$m_{\alpha 3}=y_{\alpha 3}^{\rm eff}\langle h\rangle\sim 100\; \text{GeV}$.
The remaining Majorana mass components are also fixed as
\begin{align}
	M_{1j}= M_{j1}=& \lambda_{1j}^{\rm eff}v_{\rm BL}\sim\epsilon v_{\rm BL}\sim 100\;\text{GeV},\crcr
	M_{ij}=& \lambda_{ij}^{\rm eff}v_{\rm BL}\sim v_{\rm BL}\sim 10^{13}\;\text{GeV},
\end{align}
for $i,j=(2,3)$.
To summarize, the mass matrix \eqref{eqn:MM} is written as
\begin{align}
\left(
\begin{array}{cc}
M_{\nu}^{4\times 4} & m_D^{4\times 2}\\
 & \\
(m_D^{4\times 2})^T & M^{2\times 2}
\end{array}
\right),
\end{align}
where
\begin{align}
	M_\nu^{4\times 4}\lesssim 1\; \text{eV},\\
	m_D^{4\times 2}\sim 100\; \text{GeV},\\
	M^{2\times 2}\sim 10^{13}\; \text{GeV}.
\end{align}
The seesaw formula yields the masses of the three active and one sterile neutrinos:
\begin{align}
	m_\nu^{4\times 4}\approx M_\nu^{4\times 4}-m_D^{4\times 2}(M^{2\times 2})^{-1}(m_D^{4\times 2})^T\sim 1\;\text{eV},
\end{align}
which is the desired result.
To be more precise, as the sterile neutrino mass needs to be ${\C O}(\text{eV})$ and the active neutrino masses are $\lesssim 0.1$ eV, we need $M_\nu^{4\times 4}\sim {\C O}(\text{eV})$ and  
$m_D^{4\times 2}(M^{2\times 2})^{-1}(m_D^{4\times 2})^T\lesssim 0.1$ eV in this seesaw formula.
This can be achieved by considering small ($\lesssim {\C O}(10)$) hierarchies among the Majorana and Dirac Yukawa elements.
 
\section{Final remarks} 

In this paper, we have proposed a simple scenario of warped compactification that naturally realizes large mass hierarchies of singlet neutrinos. 
With the assumptions that the right-handed neutrinos have their origin in the extra dimension and that there exist some fluctuations in the mass parameters of the bulk fermions, we have shown that the eV-scale sterile neutrino can naturally arise in the four-dimensional effective theory.
We conclude with a couple of comments.

First, we have been cavalier about the fine structure of the light neutrino masses, since such mass differences are extremely small compared to the large hierarchies that we focus on.
This certainly does not mean that they are unimportant.
See, e.g., \cite{Asaka:2011pb} for mixing between the active and sterile neutrinos.

We assumed that one of the mass parameters satisfies $\nu_1=\frac{m_1}{k}<-\half$, while the others satisfy
$\nu_2=\frac{m_2}{k}>-\half$ and $\nu_3=\frac{m_3}{k}>-\half$.
These mild conditions led to the neutrino mass hierarchies in the four-dimensional theory, with the extremely small parameter
$\epsilon=\omega_1\approx e^{-(\half+\nu_1)\pi r_c k}\sim 10^{-11}$.
The mass parameters $\nu_i$ are otherwise unconstrained.
One may, however, suppose, for example, that the $U(1)_{\rm B-L}$ symmetry breaking takes place near the Planck scale $M_4\sim M_5\sim k$; then the warp factor is fixed to be $e^{-\pi r_c k}=v_{\rm BL}/\widetilde v_{\rm BL}\sim 3\times 10^{-6}$.
In this case, one of the mass parameters is determined as $\nu_1\sim -\frac 32$.  

The original Randall-Sundrum scenario \cite{Randall:1999ee} addresses the hierarchy problem of the SM, namely, the fine-tuning issue of the Higgs potential against radiative corrections.
In our scenario, this issue of course persists, and in order to overcome this one may, for example, resort to supersymmetry.
In fact, supersymmetric extension of the scenario presented here is straightforward. 

Finally, it is now widely known that the simple 3+1 neutrino model has some tension with the global fit of short-baseline neutrino oscillation anomalies.
The purpose of our work here has been to present a minimalistic warped compactification scenario that realizes the type I seesaw mechanism with a sterile neutrino of eV mass scale.
We hope our scenario serves as a simple benchmark model for further developing more sophisticated neutrino models.


\begin{acknowledgments}
We acknowledge helpful conversations with Sudip Jana, Digesh Raut, and Carsten Rott.
We are grateful for the hospitality of the Maryland Center for Fundamental Physics, University of Maryland where this work was initiated.
This research was supported in part by the National Research Foundation of Korea Grant-in-Aid for Scientific Research No. NRF-2018R1D1A1B07051127 (S.K.) and by the U.S. Department of Energy Grant No. DE-SC0012447 (N.O.).
\end{acknowledgments} 


%
 

\end{document}